# The Leiden Ranking 2011/2012:
# Data collection, indicators, and interpretation

Ludo Waltman, Clara Calero-Medina, Joost Kosten, Ed C.M. Noyons, Robert J.W. Tijssen, Nees Jan van Eck, Thed N. van Leeuwen, Anthony F.J. van Raan, Martijn S. Visser, and Paul Wouters

Centre for Science and Technology Studies, Leiden University, The Netherlands
waltmanlr@cwts.leidenuniv.nl

The Leiden Ranking 2011/2012 is a ranking of universities based on bibliometric indicators of publication output, citation impact, and scientific collaboration. The ranking includes 500 major universities from 41 different countries. This paper provides an extensive discussion of the Leiden Ranking 2011/2012. The ranking is compared with other global university rankings, in particular the Academic Ranking of World Universities (commonly known as the Shanghai Ranking) and the Times Higher Education World University Rankings. Also, a detailed description is offered of the data collection methodology of the Leiden Ranking 2011/2012 and of the indicators used in the ranking. Various innovations in the Leiden Ranking 2011/2012 are presented. These innovations include (1) an indicator based on counting a university's highly cited publications, (2) indicators based on fractional rather than full counting of collaborative publications, (3) the possibility of excluding non-English language publications, and (4) the use of stability intervals. Finally, some comments are made on the interpretation of the ranking, and a number of limitations of the ranking are pointed out.

## 1. Introduction

The Leiden Ranking is a global university ranking based exclusively on bibliometric data. In this paper, we introduce the 2011/2012 edition of the Leiden Ranking. The paper provides a detailed discussion of the data collection methodology, the indicators used in the ranking, and the interpretation of the ranking. The Leiden Ranking 2011/2012 is available on the website www.leidenranking.com.

University rankings have quickly gained popularity, especially since the launch of Academic Ranking of World Universities, also known as the Shanghai Ranking, in 2003, and these rankings nowadays play a significant role in university decision making (Hazelkorn, 2007, 2008). The increased use of university rankings has not been hampered by the methodological problems that were already identified in an early stage (e.g., Van Raan, 2005). There are now many rankings in which universities are compared on one or more dimensions of their performance (Usher & Savino, 2006). Many of these rankings have a national or regional focus, or they consider only specific scientific disciplines. There is a small group of global university rankings (Aguillo, Bar-Ilan, Levene, & Ortega, 2010; Butler, 2010; Rauhvargers, 2011). The Leiden Ranking belongs to this group of rankings.

Global university rankings are used for a variety of purposes by different user groups. Three ways of using university rankings seem to be dominant. First, governments, funding agencies, and the media use university rankings as a source of strategic information on the global competition among universities. Second, university managers use university rankings as a marketing and decision support tool. And third, students and their parents use university rankings as a selection instrument.

An important methodological problem of the most commonly used global university rankings is their combination of multiple dimensions of university performance in a single aggregate indicator. These dimensions, which often relate to very different aspects of university performance (e.g., scientific performance and teaching performance), are combined in a quite arbitrary fashion. This prevents a clear interpretation of the aggregate indicator. A second related problem has to do with the fact that different universities may have quite different missions. Two universities that each have an excellent performance on the dimension that is most relevant to their mission may end up at very different positions in a ranking if the different dimensions are weighted differently in the aggregate indicator. These methodological problems can partly be solved by providing separate scores on the various dimensions and refraining from aggregating these scores in a single number. A third problem is more practical. Some rankings rely heavily on data supplied by the universities themselves, for instance data on staff numbers or student/staff ratios. This dependence on the universities makes these rankings vulnerable to manipulation. Also, because of the lack of internationally standardized definitions, it is often unclear to what extent data obtained from universities can be used to make valid comparisons across universities or countries.

A solution to these fundamental methodological problems is to restrict a ranking to a single dimension of university performance that can be measured in an accurate and reliable way. This is the solution that the Leiden Ranking offers. The Leiden Ranking does not attempt to measure all relevant dimensions of university performance. Instead, the ranking restricts itself to the dimension of scientific performance. Other dimensions of university performance, in particular the dimension of teaching performance, are not considered. The Leiden Ranking includes 500 major universities worldwide and is based on bibliometric data from the Web of Science database. No data is employed that has been supplied by the universities themselves. A sophisticated procedure for assigning publications to universities is used to further improve the quality of the bibliometric data.

The first edition of the Leiden Ranking was produced in 2007. In this paper, we discuss the 2011/2012 edition of the Leiden Ranking. This edition was published in December 2011 on www.leidenranking.com. Compared with earlier editions of the Leiden Ranking, the 2011/2012 edition offers a number of innovations. These innovations address some of the shortcomings of earlier editions of the ranking and also of other university rankings. Below, we summarize the most important innovations:

- The $PP_{top\ 10\%}$ indicator has been added to the Leiden Ranking. Compared with other citation impact indicators, an important advantage of the $PP_{top\ 10\%}$ indicator is its insensitivity to extremely highly cited publications.
- The fractional counting method has been added to the Leiden Ranking. We argue that, compared with the more traditional full counting method, the fractional counting method leads to more accurate comparisons between universities.
- The possibility of excluding non-English language publications has been added to the Leiden Ranking. These publications may disadvantage universities from, for instance, France and Germany.
- Stability intervals have been added to the Leiden Ranking. A stability interval provides insight into the sensitivity of an indicator to changes in the underlying set of publications.

The above innovations are discussed in more detail later on in this paper.

The rest of the paper is organized as follows. Section 2 compares the Leiden Ranking with other global university rankings. Sections 3 and 4 describe the data collection methodology of the Leiden Ranking and the indicators that are used in the ranking. Section 4 also discusses the innovations that have been made in the 2011/2012 edition of the ranking. Section 5 comments on the interpretation of the Leiden Ranking. Special attention is paid to the limitations that should be taken into account. Section 6 concludes the paper and discusses our future plans for the Leiden Ranking.

## 2. Comparison with other university rankings

Before discussing the Leiden Ranking in more detail, we compare the basic design of the ranking with the two most commonly used global university rankings: The Academic Ranking of World Universities and Times Higher Education World University Rankings. We also make a comparison with the Scimago Institutions Rankings. Like the Leiden Ranking, this is a bibliometrics-based ranking that focuses exclusively on scientific performance.

**2.1. Academic Ranking of World Universities**

The Academic Ranking of World Universities (ARWU; www.arwu.org), commonly known as the Shanghai ranking, was first published in 2003 by Shanghai Jiao Tong University (Liu & Cheng, 2005). Nowadays, the ranking is published by a company named ShanghaiRanking Consultancy. The ARWU ranking combines four criteria: Quality of education, quality of faculty, research output, and per capita performance. These criteria are measured using the following six indicators:
1. Alumni of a university winning a Nobel Prize or a Fields Medal.
2. Staff of a university winning a Nobel Prize or a Fields Medal. (Staff must be affiliated with the university at the time the prize was awarded.)
3. Highly cited researchers in 21 broad scientific fields.
4. Publications in *Nature* and *Science*.
5. Publications indexed in Web of Science (or more specifically, in the Science Citation Index Expanded and the Social Sciences Citation Index).
6. Per capita academic performance of a university. (The above five indicators normalized for a university's number of academic staff.)

The methodology of the ARWU ranking has been widely criticized. An early critical paper on the ARWU ranking was written by Van Raan (2005a; see Liu, Cheng, & Liu, 2005 and Van Raan, 2005b for the ensuing discussion). Other criticism was given by Billaut, Bouyssou, and Vincke (2010), Dehon, McCathie, and Verardi (2010), Florian (2007), Ioannidis et al. (2007), Saisana, d'Hombres, and Saltelli (2011), and Zitt and Filliatreau (2007). Examples of criticism on the ARWU ranking include the following issues:
- The weights of the six indicators are arbitrary.
- The indicators based on Nobel Prizes and Fields Medals reflect past rather than current performance and disadvantage recently established universities.
- Looking only at Nobel Prizes and Fields Medals disadvantages fields that do not have these prizes.
- Linking Nobel Prize and Fields Medal winners to universities is problematic (Enserink, 2007).
- Looking only at *Nature* and *Science* as top journals favors some fields over others and does not take into account other high quality publication venues available in many fields.
- The per capita performance indicator depends on staff numbers that may not be comparable across universities or countries.

- The ARWU ranking mainly reflects the size of a university ("Big is (made) beautiful"; Zitt & Filliatreau, 2007).

There are a number of fundamental differences between the ARWU ranking and the Leiden Ranking. First, the Leiden Ranking does not combine multiple dimensions of university performance in a single aggregate indicator. Instead, the Leiden Ranking focuses exclusively on the dimension of scientific performance. Second, the Leiden Ranking uses indicators that have been normalized for field differences. Because of this, the Leiden Ranking does not suffer from biases in favor of particular fields. Third, unlike the Nobel Prize indicators used in the ARWU ranking, the citation impact indicators used in the Leiden Ranking are based on recent data and therefore reflect the current rather than the past performance of a university. Finally, fourth, the Leiden Ranking does not rely on data supplied by the universities themselves, such as data on staff numbers.

**2.2. Times Higher Education World University Rankings**

A second well-known global university ranking is the Times Higher Education World University Rankings (Baty, 2011; www.timeshighereducation.co.uk/world-university-rankings/). We refer to this ranking simply as the THE ranking. An important element of the THE ranking is a large-scale reputational survey. In the 2011/2012 edition of the THE ranking, about 17,500 academics worldwide participated in this survey. The THE ranking combines no less than 13 indicators, categorized into five areas: Teaching, research, citations, industry income, and international outlook. Much of the data on which the THE ranking is based has been supplied by the universities themselves. Bibliometric data is taken from Web of Science. The THE ranking includes a citation impact indicator that normalizes for differences in citation behavior between scientific fields. However, the exact normalization procedure is not documented. Two other indicators, the number of PhDs awarded and the amount of research income, also include a normalization for field differences. Again, the exact normalization procedure is not clear, but data obtained from the universities seems to play a crucial role in the normalizations.

Like the ARWU ranking, the THE ranking suffers from the problem of combining multiple dimensions of university performance in a single aggregate indicator. Another problem of the THE ranking is its heavy dependence on data supplied by universities. It is unclear to what extent this data has been properly standardized and to what extent it may have been manipulated by universities. The dependence on data obtained from universities poses a clear threat to the validity of the THE ranking. The THE ranking stands out because of its reputational survey. The producers of the THE ranking consider this survey as one of the main strengths of their ranking. However, the survey has important weaknesses. Most academics know the inner workings and the real quality of only a few universities. Their impression of the vast majority of universities is based mainly on the public images of these universities and on hearsay. Rankings themselves are an important feeder of these public images. Hence, rankings such as THE may create a positive feedback loop in which well-known universities have a strong head start compared with lesser-known universities. In addition to the above issues, various other aspects of the THE ranking have been criticized as well. We refer to Bookstein, Seidler, Fieder, and Winckler (2010), Ioannidis et al. (2007), and Saisana et al. (2011) for some critical perspectives on the THE ranking.

The Leiden Ranking differs from the THE ranking in its exclusive focus on the scientific performance of universities. Moreover, unlike the THE ranking, the Leiden Ranking does not rely on questionable survey data or on data supplied by the universities themselves. There also seem to be considerable differences between the citation impact indicators used in the two rankings. However, because of the incomplete documentation of the THE ranking, the exact differences are not clear.

**2.3. Scimago Institutions Rankings**

Another global ranking, referred to as the Scimago Institutions Rankings (SIR), is produced by the Scimago research group in Spain (www.scimagoir.com). In addition to universities, the SIR ranking also includes other types of research institutions. Compared with the ARWU and THE rankings, the SIR ranking is more similar to the Leiden Ranking. Both the SIR ranking and the Leiden Ranking rely exclusively on bibliometric data, and both rankings focus on the scientific performance of institutions. Other performance dimensions are not taken into account. Despite these similarities, there are also a number of substantial methodological differences between the SIR ranking and the Leiden Ranking. The SIR ranking is based on the Scopus database, while the Leiden Ranking uses Web of Science. Because Scopus and Web of Science employ different classifications of scientific fields, this means that the two rankings have different ways of normalizing for field differences. Another difference is that the SIR ranking includes a much larger number of institutions than the Leiden Ranking (over 3000 vs. 500) and does not limit itself to universities. Unfortunately, the procedure used in the SIR ranking to identify the publications of an institution has not been documented in detail. There are also differences between the two rankings in the types of publications that are included and the indicators that are provided. Unlike the SIR ranking, the Leiden Ranking excludes arts and humanities publications, considers only a limited number of document types (i.e., *articles*, *letters*, and *reviews*), and by default does not take into account non-English language publications. Furthermore, the Leiden Ranking offers both indicators calculated using the full counting method and indicators calculated using the fractional counting method (see Subsection 4.3). In general, we consider the fractional counting method preferable over the full counting method used in the SIR ranking. Also, the Leiden Ranking offers advanced distance-based collaboration indicators (see Subsection 4.2). The SIR ranking, on the other hand, provides indicators based on the journals in which an institution has published and indicators of the degree of specialization of an institution.

**2.4. Other rankings**

We have compared the Leiden Ranking with three other global university rankings: The ARWU ranking, the THE ranking, and the SIR ranking. There are a number of global university rankings that we have not covered in our comparison. A well-known one is the QS World University Rankings (www.topuniversities.com/university-rankings/world-university-rankings/). The design of this ranking is fairly similar to the THE ranking. Another ranking that we have not covered is the Webometric Ranking of World Universities (Aguillo, Ortega, & Fernández, 2008; www.webometrics.info). The special feature of this ranking is that it is entirely based on webometric indicators. We refer to Rauhvargers (2011) for a discussion of some other global university rankings.

## 3. Data collection

In this section, we discuss the data collection methodology of the Leiden Ranking. As already mentioned, the Leiden Ranking limits itself to universities only. Other types of research institutions are not considered. Data on the publications of universities was collected from Thomson Reuters' Web of Science (WoS) database. We only considered publications of the document types *article*, *letter*, and *review* that were published between 2005 and 2009. Also, we only included publications from the sciences and the social sciences. Publications with an arts and humanities classification in WoS were excluded. Our focus was on universities with at least 500 publications in each of the five years. Changes in the organizational structure of universities were taken into consideration up to 2009. Mergers, split-ups, and other changes that took place after 2009 may not have been taken into account. Publications were assigned to universities on the basis of the institutional affiliations of authors as mentioned in the address list. The procedure for assigning publications to universities consists of two rounds, which are discussed below. We note that for most universities the results of the data collection have not been verified by the university itself.

In the first round of the publication assignment procedure, publications with the name of a university mentioned explicitly in the address list were identified. Name variants and abbreviations were taken into account as well. For instance, *Ruprecht Karls University* is a name variant of *Heidelberg University*, *TUM* of *Technische Universität München*, and *Université Paris 06* of *University Pierre and Marie Curie*. In addition, important university institutes that are mentioned in the address list of a publication without mentioning the name of the university to which they belong were assigned to the correct university. Examples include *National Environmental Research Institute of Denmark (NERI)*, which was assigned to *Aarhus University*, and *Niels Bohr Institute*, which was assigned to *University of Copenhagen*. In the first round of the publication assignment procedure, all name variants occurring at least five times were taken into account.

A key challenge in identifying the publications of a university is the way in which publications originating from academic hospitals are handled. Many medical researchers are employed by a university but actually work in an academic hospital. These researchers do not always mention their university affiliation in their publications. They sometimes provide only their contact details at the hospital. As a consequence, the publications of these researchers may not be correctly assigned to a university. At the same time, the relationship between universities and academic hospitals differs widely from one national academic system to another. In some cases, academic hospitals are an integral part of a university. In other cases, they are autonomous organizations that may collaborate with one or more universities in varying degrees and modalities. In order to prevent such differences between academic systems from having too much effect on international comparisons, there is a second round of assigning publications to universities.

In this second round, some of the publications from academic hospitals were assigned to universities. This was done on the basis of an author analysis. A publication from an academic hospital was assigned to a university if one or more authors of the publication exhibit a strong collaboration link with the university (even though the name of the university is not explicitly mentioned in the publication). An author is considered to have a strong collaboration link with a university if the name of the university is mentioned in at least half of the author's publications. As a result, for instance, some of the publications with the address *Addenbrookes Hospital* were assigned to *University of Cambridge*. Also, some of the publications with the address *Hospital La Pitié Salpêtrière* were assigned to *Paris Descartes University*, while other publications with the same address were assigned to *University Pierre and Marie Curie*.

In addition to academic hospitals, there are some other special cases in the delineation of universities. The colleges constituting *University of London* (e.g., *University College London*) were treated as separate universities. Splitting up *University of London* was done on the basis of the department field in the WoS database. Other organizations similar to *University of London* (e.g., *National University of Ireland*) were treated in the same way. Some publications produced by *University of London* and other similar organizations do not provide information about the particular college or university to which they belong. These publications were assigned to universities on the basis of an author analysis, in much the same way as publications from academic hospitals were assigned to universities. In the case of the university systems in the US, the constituent universities (e.g., *University of California, Los Angeles* and *University of Texas at Austin*) were treated as separate universities.

Since in general the publication assignment procedure of the Leiden Ranking did not take into account name variants occurring fewer than five times, and since especially the second round of the procedure inevitably involved some inaccuracies, the assignment of publications to universities is certainly not free of errors. There are two types of errors. The first one ('false positives') consists of publications that were assigned to a particular university while in fact they do not belong to this university. The second type of error ('false negatives') consists of publications that were not assigned to a particular university while in fact they do belong to this university. We expect there to be more errors of the second type than of this first type, but we estimate that universities generally do not have more than 5% errors of the second type.

Ultimately, only the 500 universities with the largest WoS publication output were included in the Leiden Ranking. The publication output of these universities ranges from about 3,200 to 61,600 publications in the period 2005–2009. The average publication output of the universities is about 9,000 publications, and the median publication output is 6,900 publications. Together, the universities have produced 3.4 million publications in the period 2005–2009. This is 61.3% of all WoS publications in this period. The 500 universities included in the Leiden Ranking are located in 41 different countries. Table 1 lists all countries with at least five Leiden Ranking universities.

Table 1. Countries with at least five Leiden Ranking universities.

| Country | No. univ. | Country | No. univ. |
| --- | --- | --- | --- |
| United States | 127 | Sweden | 10 |
| Germany | 39 | Taiwan | 9 |
| United Kingdom | 36 | Brazil | 8 |
| China | 31 | Belgium | 7 |
| Italy | 25 | Switzerland | 7 |
| Japan | 24 | Portugal | 6 |
| Canada | 21 | Finland | 6 |
| France | 20 | Greece | 6 |
| South Korea | 18 | Israel | 6 |
| Spain | 16 | Turkey | 6 |
| Australia | 14 | Austria | 5 |
| Netherlands | 12 | | |

Compared with earlier editions of the Leiden Ranking, the Leiden Ranking 2011/2012 has a more comprehensive coverage of universities from particular countries, especially from China and South Korea. The number of Chinese universities has increased from 18 in the Leiden Ranking 2010 to 31 in the Leiden Ranking 2011/2012. The number of South Korean universities has increased from 8 to 18. Australia, Brazil, India, and Taiwan each have at least three more universities in the Leiden Ranking 2011/2012 than in the Leiden Ranking 2010.

## 4. Indicators

The Leiden Ranking provides three types of indicators: Indicators of publication output, indicators of citation impact, and indicators of scientific collaboration. Publication output is measured using the *number of publications* (P) indicator. This indicator is calculated by counting the total number of publications of a university. Publications that have the document type *letter* in WoS do not count as a full publication but count as one fourth of a publication.[1] The indicators used to measure impact and collaboration are discussed in Subsections 4.1 and 4.2.

The Leiden Ranking supports two counting methods: Full counting and fractional counting. These methods differ in the way in which collaborative publications are handled. The methods are discussed in detail in Subsection 4.3. The Leiden Ranking also offers the possibility of excluding non-English language publications from the calculation of the indicators. This possibility is discussed in Subsection 4.4. Another feature of the Leiden Ranking is the possibility to complement indicators with so-called stability intervals. This feature is discussed in Subsection 4.5.

Unless stated otherwise, the empirical results reported in this section were obtained using the fractional counting method based on English-language publications only. We also note that all results of the Leiden Ranking are available in an Excel file that can be downloaded from www.leidenranking.com.

**4.1. Impact indicators**

The Leiden Ranking includes three indicators of the citation impact of the work of a university:

- *Mean citation score* (MCS). The average number of citations of the publications of a university.
- *Mean normalized citation score* (MNCS). The average number of citations of the publications of a university, normalized for differences between scientific fields (i.e., WoS subject categories), differences between publication years, and differences between document types (i.e., *article*, *letter*, and *review*). An MNCS value of one can be interpreted as the world average (or more properly, the average of all WoS publications). Consequently, if a university has an MNCS value of two, this for instance means that the publications of the university have been cited twice above world average. We refer to Waltman, Van Eck, Van Leeuwen, Visser, and Van Raan (2011) for a more detailed discussion of the MNCS indicator.
- *Proportion top 10% publications* (PP$_{top\ 10\%}$). The proportion of the publications of a university that, compared with other similar publications, belong to the top 10% most frequently cited (Tijssen, Visser, & Van Leeuwen, 2002). Publications are considered similar if they were published in the same field and the same publication year and if they have the same document type. We note that an indicator similar to our PP$_{top\ 10\%}$ indicator was recently introduced in the Scimago Institutions Rankings (Bornmann, De Moya-Anegón, & Leydesdorff, in press).

---

[1] Counting letters as one fourth of an ordinary publication (i.e., an article or a review) of course involves some arbitrariness. We have chosen to use a weight of 0.25 for letters because in WoS a letter on average receives roughly one fourth of the citations of an ordinary publication.

In the calculation of the above indicators, citations are counted until the end of 2010. Author self citations are excluded from all calculations.[2] Publications of the document type *letter* are weighted as one fourth of a full publication.[3]

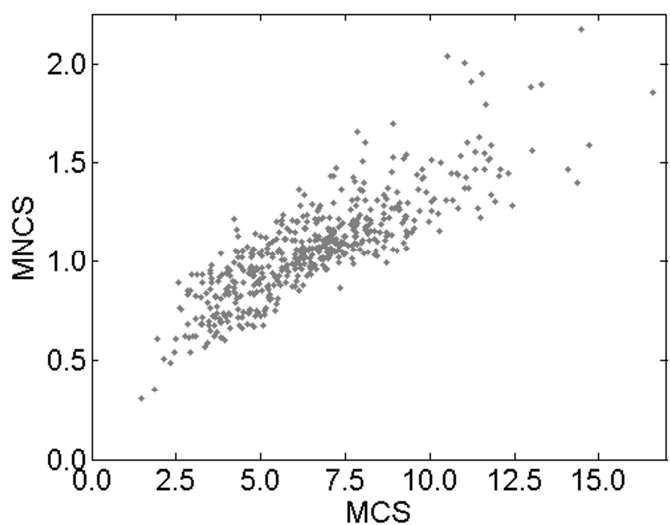

Figure 1. Scatter plot of the relation between the MCS indicator and the MNCS indicator for the 500 Leiden Ranking universities ($r = 0.84$).

Figure 1 provides a scatter plot of the relation between the MCS indicator and the MNCS indicator for the 500 Leiden Ranking universities. The two indicators are clearly correlated with each other, but their relation is not very strong. This shows that the normalization performed by the MNCS indicator has a quite significant effect on the way in which citation impact is assessed. As an example, consider *Massachusetts Institute of Technology* and *University of Massachusetts Medical School*. Based on the MCS indicator, these universities are ranked 3rd and 4th in the Leiden Ranking, with values of 14.46 and 14.37, respectively. However, the two universities have quite different scientific profiles. *University of Massachusetts Medical School* focuses strongly on medical research, while *Massachusetts Institute of Technology* is more broadly oriented, with an emphasis on natural sciences and engineering research. The fields in which *Massachusetts Institute of Technology* is active generally have a lower citation density than the fields in which *University of Massachusetts Medical School* publishes most of its research. Because of this, even though the two universities have similar MCS values, the impact of the work of *Massachusetts Institute of Technology* should be assessed considerably higher than the impact of the work of *University of Massachusetts Medical School*. This is indeed reflected by the MNCS indicator. The MNCS indicator equals 2.17 for *Massachusetts Institute of Technology* (ranked 1st), while it equals 1.40 for *University of Massachusetts Medical School* (ranked 50th).

---

[2] A citation is regarded as an author self citation if the citing and the cited publication have at least one author name (i.e., last name and initials) in common. We tested the effect of excluding self citations on the MNCS and $PP_{top\ 10\%}$ indicators. For most universities, the effect turns out to be negligible. There is a small set of universities for which the effect is more substantial. These are mainly universities from continental Europe, especially from Germany. For these universities, excluding self citations considerably decreases the MNCS and $PP_{top\ 10\%}$ indicators.

[3] The 2011/2012 edition of the Leiden Ranking uses different impact indicators than earlier editions of the ranking. The MCS and MNCS indicators in the 2011/2012 edition are comparable with the CPP and CPP/FCSm indicators in earlier editions of the ranking. We refer to Waltman et al. (2011) for a discussion of the difference between the MNCS indicator and the CPP/FCSm indicator. In earlier editions of the Leiden Ranking, no indicator similar to the $PP_{top\ 10\%}$ indicator was used.

Figure 2 shows a scatter plot of the relation between the MNCS indicator and the $PP_{top\ 10\%}$ indicator for the 500 Leiden Ranking universities. There is a strong, more or less linear relation between the two indicators. However, there is one university for which the indicators deviate strongly from this relation. This is *University of Göttingen*. This university is ranked 2nd based on the MNCS indicator, while it is ranked 238th based on the $PP_{top\ 10\%}$ indicator. The MNCS indicator for *University of Göttingen* turns out to have been strongly influenced by a single extremely highly cited publication. This publication (Sheldrick, 2008) was published in January 2008 and had been cited over 16,000 times by the end of 2010. Without this single publication, the MNCS indicator for *University of Göttingen* would have been equal to 1.09 instead of 2.04, and *University of Göttingen* would have been ranked 219th instead of 2nd. Unlike the MNCS indicator, the $PP_{top\ 10\%}$ indicator is hardly influenced by a single very highly cited publication. This is because the $PP_{top\ 10\%}$ indicator only takes into account whether a publication belongs to the top 10% of its field or not. The indicator is insensitive to the exact number of citations of a publication. This is an important difference with the MNCS indicator, and this difference explains why in the case of *University of Göttingen* the MNCS indicator and the $PP_{top\ 10\%}$ indicator yield very different results. In our view, the sensitivity of the MNCS indicator to a single very highly cited publication is an undesirable property. We therefore regard the $PP_{top\ 10\%}$ indicator as the most important impact indicator in the Leiden Ranking.

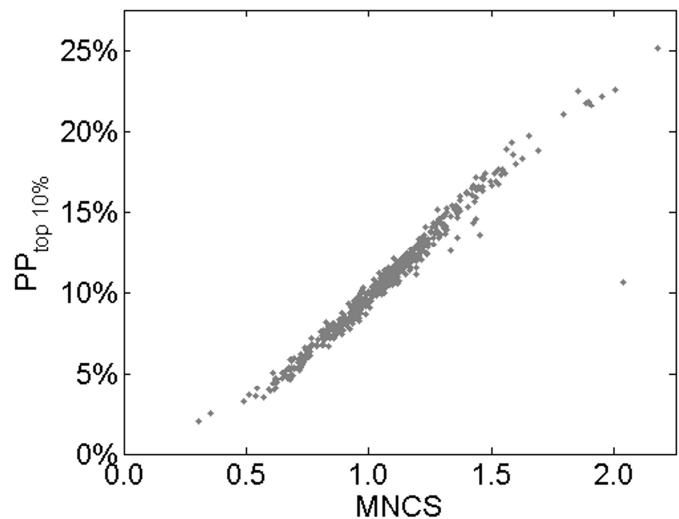

Figure 2. Scatter plot of the relation between the MNCS indicator and the $PP_{top\ 10\%}$ indicator for the 500 Leiden Ranking universities ($r = 0.98$).

The $PP_{top\ 10\%}$ indicator may be criticized because focusing exclusively on top 10% publications is somewhat arbitrary. For instance, why not use top 5% or top 20% publications? Figure 3 shows that at the level of universities the exact threshold that is used is not really important. The $PP_{top\ 5\%}$ and $PP_{top\ 20\%}$ indicators yield very similar results as the $PP_{top\ 10\%}$ indicator.

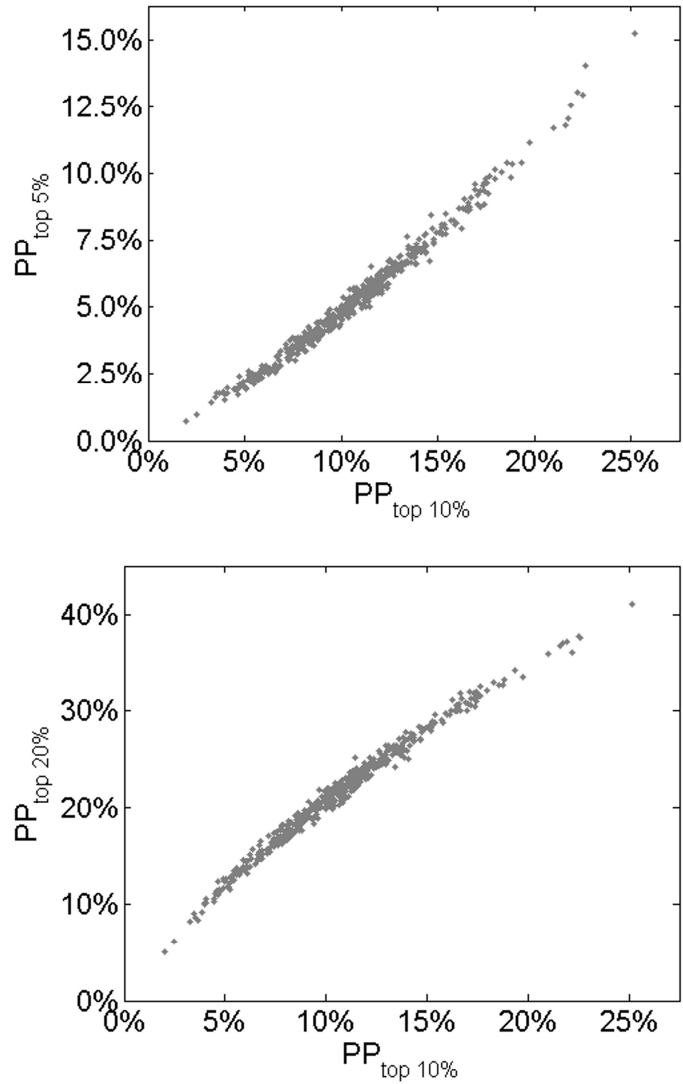

Figure 3. Scatter plot of the relation between the $PP_{top\ 10\%}$ indicator and the $PP_{top\ 5\%}$ indicator (top panel; $r = 0.99$) and the relation between the $PP_{top\ 10\%}$ indicator and the $PP_{top\ 20\%}$ indicator (bottom panel; $r = 0.99$) for the 500 Leiden Ranking universities.

**4.2. Collaboration indicators**

The Leiden Ranking includes four indicators of the degree to which a university is involved in scientific collaborations with other organizations:

- *Proportion collaborative publications* ($PP_{collab}$). The proportion of the publications of a university that have been co-authored with one or more other organizations.
- *Proportion international collaborative publications* ($PP_{int\ collab}$). The proportion of the publications of a university that have been co-authored by two or more countries.

- *Mean geographical collaboration distance* (MGCD). The average geographical collaboration distance of the publications of a university. The geographical collaboration distance of a publication is defined as the largest geographical distance between two addresses mentioned in the publication's address list. If a publication's address list contains only one address, the geographical collaboration distance of the publication equals zero. We refer to Tijssen, Waltman, and Van Eck (2011) and Waltman, Tijssen, and Van Eck (2011) for a more detailed discussion of the MGCD indicator, including a discussion of the geocoding procedure that was used to identify the geographical coordinates of the addresses mentioned in publications' address lists.[4]
- *Proportion long distance collaborative publications* ($PP_{>1000\ km}$). The proportion of the publications of a university that have a geographical collaboration distance of more than 1000 km.

Like in the impact indicators discussed in Subsection 4.1, publications of the document type *letter* are weighted as one fourth of a full publication in the above indicators.

Table 2 reports the Pearson correlations between the above four collaboration indicators. The correlations were calculated based on the indicator values of the 500 Leiden Ranking universities. As can be seen in the table, the correlations between the $PP_{collab}$ and $PP_{int\ collab}$ indicators on the one hand and the MGCD and $PP_{>1000\ km}$ indicators on the other hand are all very low. The correlations between the $PP_{collab}$ indicator and the $PP_{int\ collab}$ indicator and between the MGCD indicator and the $PP_{>1000\ km}$ indicator are somewhat higher, but still not very high. This indicates that each of the four indicators measures a different aspect of scientific collaboration.

Table 2. Pearson correlations between the four collaboration indicators included in the Leiden Ranking.

|  | $PP_{collab}$ | $PP_{int\ collab}$ | MGCD | $PP_{>1000\ km}$ |
|---|---|---|---|---|
| $PP_{collab}$ | 1.00 | 0.55 | 0.14 | 0.12 |
| $PP_{int\ collab}$ | 0.55 | 1.00 | 0.25 | 0.17 |
| MGCD | 0.14 | 0.25 | 1.00 | 0.74 |
| $PP_{>1000\ km}$ | 0.12 | 0.17 | 0.74 | 1.00 |

There are two distance-based collaboration indicators: The MGCD indicator and the $PP_{>1000\ km}$ indicator. To illustrate how these two indicators complement each other, Figure 4 shows a scatter plot of the relation between the indicators for the 500 Leiden Ranking universities. A more or less linear relation can be observed, but there are approximately 30 universities for which the indicators do not follow this linear relation. For these universities, the MGCD value is relatively high compared with the $PP_{>1000\ km}$ value. With one exception (i.e., *London School of Hygiene & Tropical Medicine*), the 30 universities all turn out to have geographically quite peripheral locations. The universities are located in Argentina, Australia, Chile, New Zealand, Singapore, South Africa, Thailand, and in Honolulu, Hawaii. This illustrates how the combination of the MGCD indicator and the $PP_{>1000\,km}$ indicator reveals the special collaboration characteristics of universities in peripheral locations.

---

[4] We note that there are some small inaccuracies in the calculation of the MGCD indicator. This is because for some of the addresses mentioned in publications' address lists we do not have the geographical coordinates. In the case of the Leiden Ranking universities, about 2.3% of the publications have at least one address without geographical coordinates. Addresses without geographical coordinates are ignored in the calculation of the MGCD indicator.

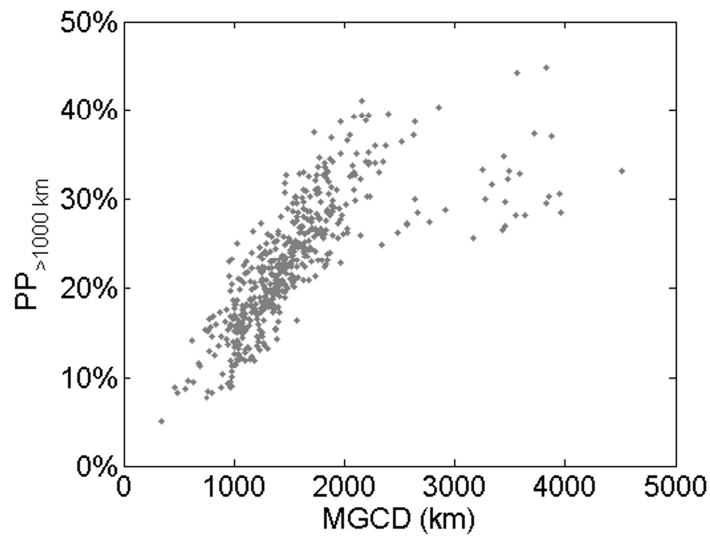

Figure 4. Scatter plot of the relation between the MGCD indicator and the $PP_{>1000\,km}$ indicator for the 500 Leiden Ranking universities ($r = 0.74$).

**4.3. Counting method**

The Leiden Ranking supports two counting methods: Full counting and fractional counting. In the calculation of the indicators, the full counting method gives equal weight to all publications of a university (except for publications of the document type *letter*). The fractional counting method gives less weight to collaborative publications than to non-collaborative ones. For instance, if the address list of a publication contains five addresses and two of these addresses belong to a particular university, then the publication has a weight of 0.4 in the calculation of the indicators for this university. Using the fractional counting method, a publication is fully assigned to a university only if all addresses mentioned in the publication's address list belong to the university.

For the purpose of making comparisons between universities, we consider the fractional counting method preferable over the full counting method. This is based on the following argument. If for each publication in WoS we calculate the MNCS indicator, the average of all these publication-level MNCS values will be equal to one. We want a similar property to hold at the level of organizations. If each publication in WoS belongs to one or more organizations and if for each organization we calculate the MNCS indicator, we want the average (weighted by publication output) of all these organization-level MNCS values to be equal to one. If this property holds, the value one can serve as a benchmark not only at the level of publications but also at the level of organizations. This would for instance mean that an organization with an MNCS indicator of two can be said to perform twice above average in comparison with other organizations. Using the full counting method, however, the above property does not hold. This is because publications belonging to multiple organizations are fully counted multiple times, once for each organization to which they belong. This double counting of publications causes the average of the organization-level MNCS values to deviate from one. Using the fractional counting method, on the other hand, it can be shown that the above property does hold. Therefore, if the fractional counting method is used, the value one can serve as a proper benchmark at the organization level. This is our main argument for preferring the fractional counting method over the full counting method.[5] We note that the argument is not restricted to the MNCS indicator. The argument also applies to other indicators, such as the $PP_{top\ 10\%}$ indicator.

In practice, the full counting method causes the average of the organization-level MNCS values to be greater than one. Similarly, it causes the average of the organization-level $PP_{top\ 10\%}$ values to be greater than 10%. This is due to a combination of two mechanisms. First, collaborative publications are counted multiple times in the full counting method, and second, collaborative publications tend to be cited more frequently than non-collaborative publications. The combination of these two mechanisms is responsible for the effect that at the level of organizations MNCS and $PP_{top\ 10\%}$ values on average are greater than, respectively, one and 10%. Importantly, there are substantial differences between scientific fields in the strength of this effect. For instance, the effect is very strong in clinical medicine and quite weak in chemistry, engineering, and mathematics.[6] Because of these differences between fields, the full counting method may be considered biased in favor of some organizations over others. Organizations active mainly in clinical medicine research for instance have an advantage over organizations focusing on engineering research.

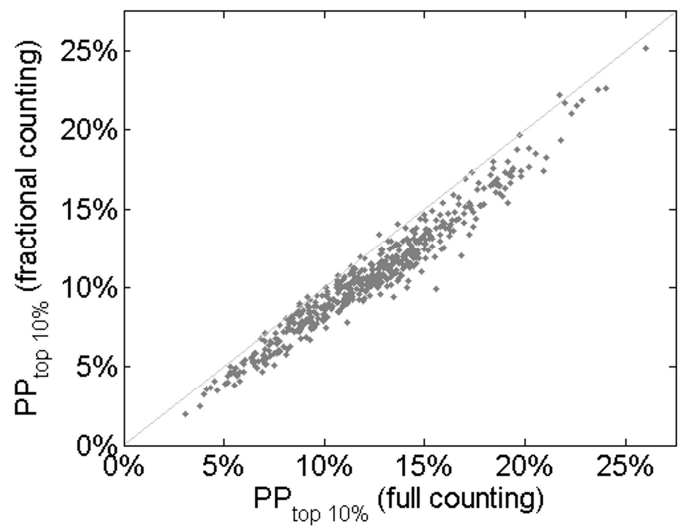

Figure 5. Scatter plot of the relation between the full counting $PP_{top\ 10\%}$ indicator and the fractional counting $PP_{top\ 10\%}$ indicator for the 500 Leiden Ranking universities ($r = 0.97$).

---

[5] A similar argument in favor of the fractional counting method is given in a recent paper by Aksnes, Schneider, and Gunnarsson (2012), in which the full counting method and the fractional counting method are compared at the level of countries. We refer to Gauffriau, Larsen, Maye, Roulin-Perriard, and Von Ins (2008) for an overview of the literature on counting methods.

[6] This statement is based on the following analysis. For different scientific fields, we calculated the average number of citations per publication. Both a weighted and an unweighted average were calculated. In the case of the weighted average, each publication was weighted by the number of addresses in the address list. The ratio of the weighted and the unweighted average provides an indication of the size of the 'full counting bonus'. A ratio of approximately 1.35 was obtained for clinical medicine. Ratios below 1.10 were obtained for chemistry, engineering, and mathematics.

Figure 5 shows a scatter plot of the relation between the $PP_{top\ 10\%}$ indicator calculated using the full counting method and the $PP_{top\ 10\%}$ indicator calculated using the fractional counting method. For almost all universities, the $PP_{top\ 10\%}$ indicator calculated using the full counting method has a higher value than the $PP_{top\ 10\%}$ indicator calculated using the fractional counting method. This is a consequence of the 'full counting bonus' discussed above. The overall correlation between the full counting method and the fractional counting method turns out to be high ($r = 0.97$), but as can be seen in Figure 5, at the level of individual universities the difference between the two counting methods can be quite significant. Tables 3 and 4 list the ten universities that, based on the $PP_{top\ 10\%}$ indicator, benefit most from either the full or the fractional counting method. In line with the above discussion, the universities benefiting from the full counting method almost all have a clear medical profile. (Exceptions are *University of Nantes* and *University of Hawaii, Mānoa*.) The other way around, the universities benefiting from the fractional counting method all have a strong focus on engineering research and on the natural sciences. Most of these universities are located in Asia.

Table 3. The ten universities that, based on the $PP_{top\ 10\%}$ indicator, benefit most from the full counting method.

| University | Country | $PP_{top\ 10\%}$ indicator | |
|---|---|---|---|
| | | Full counting | Fractional counting |
| Lille 2 University of Health and Law | France | 15.6% | 9.9% |
| Wake Forest University | United States | 16.8% | 12.0% |
| Hannover Medical School | Germany | 14.1% | 10.0% |
| University of Nantes | France | 13.5% | 9.4% |
| University of Alabama at Birmingham | United States | 14.9% | 11.0% |
| University of Colorado Denver | United States | 17.2% | 13.4% |
| Medical College of Wisconsin | United States | 14.2% | 10.4% |
| Mount Sinai School of Medicine | United States | 19.2% | 15.4% |
| Saint Louis University | United States | 14.2% | 10.4% |
| University of Hawaii, Mānoa | United States | 15.5% | 11.9% |

Table 4. The ten universities that, based on the $PP_{top\ 10\%}$ indicator, benefit most from the fractional counting method.

| University | Country | $PP_{top\ 10\%}$ indicator | |
|---|---|---|---|
| | | Full counting | Fractional counting |
| Nankai University | China | 12.7% | 13.4% |
| Rice University | United States | 21.7% | 22.2% |
| Pohang University of Science and Technology | South Korea | 13.7% | 14.1% |
| Indian Institute of Technology Kharagpur | India | 8.7% | 9.0% |
| National Chung Hsing University | Taiwan | 9.2% | 9.4% |
| Lanzhou University | China | 11.8% | 11.9% |
| Indian Institute of Technology Madras | India | 8.7% | 8.8% |
| Sichuan University | China | 7.0% | 7.1% |
| Rensselaer Polytechnic Institute | United States | 17.3% | 17.4% |
| Nanjing University | China | 10.7% | 10.7% |

### 4.4. Non-English language publications

About 2.1% of the publications of the Leiden Ranking universities have not been written in English. Of these non-English language publications, most have been written in German (31%), Chinese (17%), French (17%), Spanish (13%), or Portuguese (10%). Comparing the impact of non-English language publications with the impact of publications written in English may not be considered fair (Van Leeuwen, Moed, Tijssen, Visser, & Van Raan, 2001; Van Raan, Van Leeuwen, & Visser, 2011a, 2011b). Non-English language publications can be read only by a small part of the scientific community, and therefore these publications cannot be expected to receive similar numbers of citations as publications written in English. To deal with this issue, the Leiden Ranking offers the possibility of excluding non-English language publications from the calculation of the indicators.

Figure 6 shows a scatter plot of the relation between the $PP_{top\ 10\%}$ indicator based on all publications and the $PP_{top\ 10\%}$ indicator based on English-language publications only. The overall correlation is very high ($r = 0.99$), and for most universities including or excluding non-English language publications makes hardly any difference. Nevertheless, there are a number of universities that benefit quite significantly from excluding non-English language publications. These are mostly French and German universities, but also some from China and other countries. Table 5 lists the ten universities that, based on the $PP_{top\ 10\%}$ indicator, benefit most from excluding non-English language publications. We refer to Van Raan et al. (2011a) for additional empirical results on the effect of excluding non-English language publications.

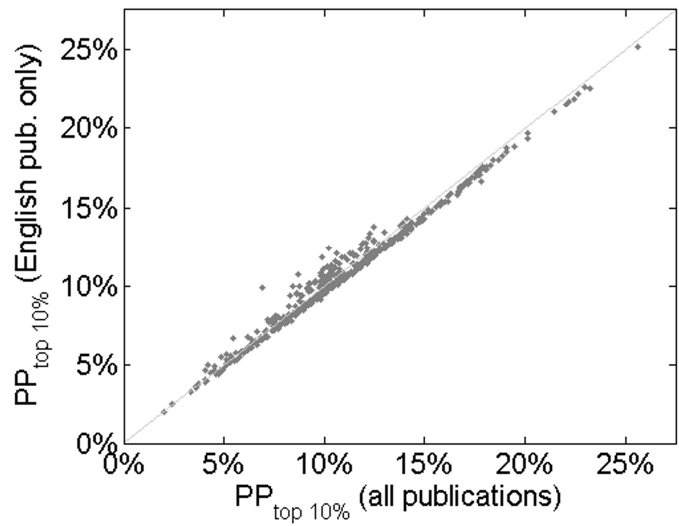

Figure 6. Scatter plot of the relation between the $PP_{top\ 10\%}$ indicator based on all publications and the $PP_{top\ 10\%}$ indicator based on English-language publications only for the 500 Leiden Ranking universities ($r = 0.99$).

Table 5. The ten universities that, based on the $PP_{top\ 10\%}$ indicator, benefit most from excluding non-English language publications.

| University | Country | $PP_{top\ 10\%}$ indicator | |
| --- | --- | --- | --- |
| | | All publications | English pub. only |
| Lille 2 University of Health and Law | France | 6.9% | 9.9% |
| Université Bordeaux Segalen | France | 10.2% | 12.5% |
| Montpellier 1 University | France | 8.7% | 10.7% |
| Paris Descartes University | France | 9.9% | 11.9% |
| Université de la Méditeranée - Aix-Marseille II | France | 8.4% | 10.0% |
| University of Nice Sophia Antipolis | France | 10.6% | 12.1% |
| Ludwig-Maximilians-Universität München | Germany | 12.5% | 13.8% |
| Pontifical Catholic University of Chile | Chile | 5.5% | 6.7% |
| Hannover Medical School | Germany | 8.8% | 10.0% |
| University of Ulm | Germany | 9.9% | 11.1% |

**4.5. Stability intervals**

The stability of an indicator relates to the sensitivity of the indicator to changes in the underlying set of publications. An indicator has a low stability if it is highly sensitive to changes in the set of publications based on which it is calculated. An indicator has a high stability if it is relatively insensitive to such changes. For instance, if a university has one or a few very highly cited publications and a large number of lowly cited publications, the MNCS indicator for this university will be relatively unstable. This is because the value of the MNCS indicator depends strongly on whether the university's highly cited publications are included in the calculation of the indicator or not. A university whose publications all have similar citation scores will have a very stable MNCS indicator. In general, the larger the number of publications of a university, the more stable the indicators calculated for the university.

To provide some insight into the stability of indicators, the Leiden Ranking uses so-called stability intervals. Stability intervals are similar to confidence intervals, but they have a somewhat different interpretation. A stability interval indicates a range of values of an indicator that are likely to be observed when the underlying set of publications changes. For instance, the MNCS indicator may be equal to 1.50 for a particular university, with a stability interval from 1.40 to 1.65. This means that the true value of the MNCS indicator equals 1.50 for this university, but that changes in the set of publications of the university may relatively easily lead to MNCS values in the range from 1.40 to 1.65. The larger the stability interval of an indicator, the lower the stability of the indicator.

The stability intervals used in the Leiden Ranking are constructed as follows. Consider a university with $n$ publications, and suppose we want to construct a stability interval for the MNCS indicator of this university. We then randomly draw 1000 samples from the set of publications of the university. Each sample is drawn with replacement, which means that a publication may occur multiple times in the same sample. The size of each sample is $n$, which is equal to the number of publications of the university. For each sample, we calculate the value of the MNCS indicator. This yields a distribution of 1000 sample MNCS values. We use this distribution to determine a stability interval for the MNCS indicator of the university. The Leiden Ranking uses 95% stability intervals. To obtain a 95% stability interval, we take the 2.5th and the 97.5th percentile of the distribution of sample MNCS values. These percentiles serve as the lower and the upper bound of the stability interval.

In the statistical literature, the above procedure for constructing stability intervals is known as bootstrapping (Efron & Tibshirani, 1993; Spiegelhalter & Goldstein, 2009). Stability intervals are also discussed in a recent paper by Colliander and Ahlgren (2011). However, Colliander and Ahlgren use a somewhat different procedure for constructing stability intervals than we do.[7]

Figure 7 shows the MNCS indicator and the corresponding stability interval for 50 Leiden Ranking universities. These are the 50 universities that are ranked highest based on the MNCS indicator. Most of the universities have a small stability interval (sometimes almost invisible in Figure 7). This is not really surprising. All universities in the Leiden Ranking have a quite large publication output, and in general a large publication output leads to small stability intervals. However, there are two universities with a remarkably large stability interval. One is *University of Göttingen* (ranked 2nd), and the other is *Utrecht University* (ranked 35th). *University of Göttingen* has a stability interval that ranges from 1.06 to 3.95. The stability interval of *Utrecht University* is smaller, but its range from 1.24 to 1.86 is still rather large. In both cases, the MNCS indicator turns out to have been strongly influenced by a single publication with a very large number of citations. (For *University of Göttingen*, this was already observed in Subsection 4.1.) The MNCS indicator is highly sensitive to publications with a very large number of citations. These publications strongly reduce the stability of the MNCS indicator, and this leads to large stability intervals.

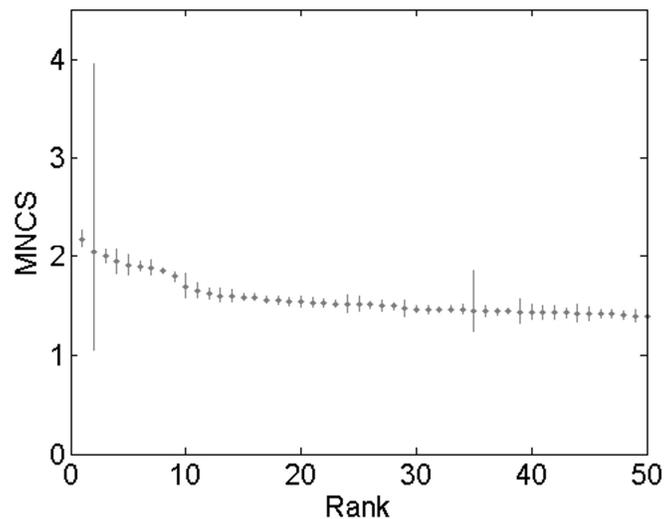

Figure 7. MNCS indicator and the corresponding stability interval for the 50 Leiden Ranking universities with the highest MNCS-based ranking.

---

[7] Leydesdorff and Bornmann (in press) suggest the use of a statistical test to determine whether differences between universities in the Leiden Ranking are statistically significant. This can be seen as an alternative to our stability intervals. However, given the various problems associated with statistical tests (Schneider, 2011), we prefer the use of our stability intervals.

## 5. Interpretation of the Leiden Ranking

University rankings aim to capture a complex reality in a small set of numbers. By necessity, this imposes serious limitations on these rankings. Below, we summarize a number of important limitations that should be taken into account in the interpretation of the Leiden Ranking:

1. The Leiden Ranking does not capture the teaching performance of universities. Instead, the Leiden Ranking focuses exclusively on universities' scientific performance, and the scientific performance of a university need not be a good predictor of its teaching performance. In addition, only specific aspects of the scientific performance of a university are taken into account in the Leiden Ranking, in particular publication output and citation impact in WoS covered journals. Other aspects of a university's scientific performance, such as its impact in national journals (not covered in WoS) or the societal impact of its research, are not considered in the Leiden Ranking.
2. The Leiden Ranking is based exclusively on output variables of the process of scientific research (i.e., publications, citations, and co-authorships). Input variables, such as the number of research staff of a university or the amount of money a university has available for research, are not taken into account. Ideally, scientific performance should be measured based on both input and output variables (see also Calero-Medina, López-Illescas, Visser, & Moed, 2008). However, accurate internationally standardized data on input variables is not available, and this is why the Leiden Ranking uses output variables only.
3. Indicators like those used in the Leiden Ranking (but also in other university rankings) can be quite sensitive to all kinds of choices regarding the details of their calculation. This is well illustrated by the empirical results presented in Sections 4.3 and 4.4, which show the effect of the choice of a counting method and the effect of the choice to include or exclude non-English language publications. The sensitivity of indicators to choices like these should be kept in mind in the interpretation of the Leiden Ranking. Many choices are somewhat hidden in the details of the calculation of an indicator and may appear to be of a rather technical nature. Nevertheless, these choices may significantly affect the results produced by an indicator. In a sense, what this means is that the results produced by an indicator are always subject to a certain degree of uncertainty. If the indicator had been calculated in a slightly different way, the results would have looked differently. It is important to emphasize that this type of uncertainty is difficult to quantify and is not reflected in the stability intervals discussed in Section 4.5. Stability intervals only reflect uncertainty related to changes in the set of publications underlying an indicator.
4. In the interpretation of university rankings, attention often focuses almost completely on the ranks of universities (e.g., "University X is ranked 20 positions higher than university Y" or "Country Z has five universities in the top 100"). This type of interpretation has the advantage of being easy to understand by a broad audience. However, the interpretation can also be misleading. Using indicators such as MNCS or $PP_{top\ 10\%}$, the performance of universities tends to be quite skewed. There are a small number of universities with a very high performance (e.g., MNCS above 1.8) and a large number of universities with a more average performance (e.g., MNCS between 1.0 and 1.5). This for instance means that according to the MNCS indicator the difference between the universities on ranks 1 and 10 in the Leiden Ranking is almost 0.5 while the difference between the universities on ranks 200 and 300 is less than 0.1. In other words, an increase in the rank of a university by, say, ten positions is much more significant in the top of the ranking than further down the list. A more accurate interpretation of university rankings in general and of the Leiden Ranking in particular can be obtained by looking directly at the values of the indicators rather than at the rankings implied by these values. For instance, "University X is performing 20% better than university Y" is more accurate than "University X is ranked 20 positions higher than university Y".

5. The Leiden Ranking assesses universities as a whole and therefore cannot be used to draw conclusions regarding the performance of individual research groups, departments, or institutes within a university. Different units within the same university may differ quite a lot in their performance, and drawing conclusions at the level of individual units based on the overall performance of a university is therefore not allowed (see also López-Illescas, De Moya-Anegón, & Moed, 2011). More detailed bibliometric analyses are needed to draw conclusions at the level of individual units within a university.

## 6. Conclusion

In this paper, we have introduced the Leiden Ranking 2011/2012. A detailed discussion has been provided of the data collection methodology, the indicators used in the ranking, and the interpretation of the ranking.

Compared with other global university rankings, in particular the popular ARWU and THE rankings, the Leiden Ranking offers a number of important advantages. First, the Leiden Ranking refrains from arbitrarily combining multiple dimensions of university performance in a single aggregate indicator. Second, the Leiden Ranking does not rely on data supplied by the universities themselves and also does not use questionable survey data. And third, the Leiden Ranking is extensively documented, making it more transparent than many other rankings.

At the same time, we also acknowledge a number of limitations of the Leiden Ranking. Depending on the purpose for which a university ranking is used, the exclusive focus of the Leiden Ranking on scientific performance can be a serious limitation. Because of this limitation, the Leiden Ranking is not very useful for prospective undergraduate students in their choice of a university. The Leiden Ranking captures the scientific performance of a university mainly by measuring the citation impact of the university's publications. This also involves some limitations. On the one hand, citation impact is only one element of scientific performance. It does not capture elements such as the societal impact of the work of a university. On the other hand, the measurement of citation impact has various methodological difficulties, for instance because of restrictions imposed by the Web of Science database, because of limitations of the indicators that are used, and because of intrinsic difficulties associated with certain scholarly disciplines (e.g., humanities and some of the social sciences). Another shortcoming of the Leiden Ranking is the absence of a disciplinary breakdown. The Leiden Ranking offers statistics only at the level of science as a whole. Clearly, for many purposes, more fine-grained statistics are needed, for instance at the level of individual scientific fields. Such statistics are not available in the Leiden Ranking, but they can be calculated as part of performance analyses for specific universities.

We plan to further extend the Leiden Ranking in the next editions. We are considering extensions in three directions. First, the number of universities included in the Leiden Ranking may be increased, and other types of research institutions may be added to the ranking. Also, a classification of universities (e.g., 'general university', 'medical university', 'technical university', etc.) may be developed in order to facilitate comparisons among similar entities. Second, the statistics offered by the Leiden Ranking may be refined, for instance by reporting longitudinal trends and by providing a breakdown into a number of broad scientific disciplines. And third, the indicators used in the Leiden Ranking may be improved, and new indicators may be added. For instance, there may be room for a more sophisticated approach to the normalization of impact indicators for field differences, and an indicator of university-industry collaboration (Tijssen, 2012) may be added to the ranking.

Some of the above innovations are likely to spill over to developments outside the Leiden Ranking, in particular to U-Multirank (www.u-multirank.eu). U-Multirank is a new user-driven interactive tool for classifying, benchmarking, and ranking of universities worldwide. Our institute is involved in the development of this tool. U-Multirank will offer a multidimensional ranking of universities that includes indicators on a variety of dimensions of university performance (i.e., teaching and learning, research, knowledge transfer, international orientation, and regional engagement). Unlike many existing university rankings, U-Multirank will not combine these indicators in a single aggregate indicator. We refer to Van Vught and Westerheijden (2010) and Van Vught and Ziegele (2012) for more details on U-Multirank.

## Acknowledgment

We would like to thank Henk Moed for his contribution to earlier editions of the Leiden Ranking.## References

Aguillo, I.F., Ortega, J.L., & Fernández, M. (2008). Webometric Ranking of World Universities: Introduction, methodology, and future developments. *Higher Education in Europe*, *33*(2–3), 233–244.
Aguillo, I.F., Bar-Ilan, J., Levene, M., & Ortega, J.L. (2010). Comparing university rankings. *Scientometrics*, *85*(1), 243–256.
Aksnes, D.W., Schneider, J.W., & Gunnarsson, M. (2012). Ranking national research systems by citation indicators. A comparative analysis using whole and fractionalised counting methods. *Journal of Informetrics*, *6*(1), 36–43.
Baty, P. (2011). *Change for the better*. Retrieved February 7, 2012, from http://www.timeshighereducation.co.uk/world-university-rankings/2011-2012/analysis-rankings-methodology.html.
Billaut, J.-C., Bouyssou, D., & Vincke, P. (2010). Should you believe in the Shanghai ranking? An MCDM view. *Scientometrics*, *84*(1), 237–263.
Bookstein, F.L., Seidler, H., Fieder, M., & Winckler, G. (2010). Too much noise in the Times Higher Education rankings. *Scientometrics*, *85*(1), 295–299.
Bornmann, L., De Moya-Anegón, F., & Leydesdorff, L. (in press). The new excellence indicator in the World Report of the SCImago Institutions Rankings 2011. *Journal of Informetrics*.
Butler, D. (2010). University rankings smarten up. *Nature*, *464*, 16–17.21